
\input amstex
\magnification=\magstep1
\vsize=23truecm
\hsize=15.5truecm
\hoffset=2.4truecm
\voffset=1.5truecm
\parskip=.2truecm
\font\ut=cmbx10
\font\ti=cmbx10 scaled\magstep1

\font\ninerm=cmr9

\def\al{\alpha}
\def\be{\beta}
\def\ab{{\alpha\beta}}
\def\de{\delta}
\def\ga{\gamma}
\def\la{\lambda}
\def\hg{{\hat g}}
\def\pa{\partial}

\def\pr{\prime}
\def\si{\sigma}

\font\ut=cmbx10 scaled\magstep1
\font\ti=cmbx10 scaled\magstep2

%
\baselineskip=.6truecm
{\hfill ZU-TH-4/92}
\vskip.1truecm
{\hfill March 1992}
\vskip2.5truecm
\centerline{\ti Classical and Quantum Considerations}
\vskip0.1truecm
\centerline{\ti of Two-dimensional Gravity}
\vskip0.3truecm
\centerline{\ }
\vskip2.0truecm
\centerline{
T. T. Burwick\,
\plainfootnote{$^{a)}$}
{\ninerm Supported by the Swiss National Science Foundation}
\plainfootnote{$^{b)}$}
{\ninerm Adress after 1 may 1992: SLAC,
Stanford University, Stanford, CA 94309, USA}
and\,A. H. Chamseddine$\, ^{a)}$
}
\vskip1truecm
\centerline{\it Institute for Theoretical Physics}
\centerline{\it University of Z\"urich}
\centerline{\it Sch\"onberggasse 9}
\centerline{\it CH-8001 Z\"urich}
\centerline{\it Switzerland}
\vskip2.5truecm
\centerline{\bf Abstract}
{\sl
The two-dimensional theory of gravity describing a
graviton-dilaton system is considered.
The graviton-dilaton coupling can be
fixed such that the quantum theory remains free of the conformal anomaly
for any conformal dimension of the coupled matter system,
even if the dilaton
does not appear
as Lagrange multiplier.
Interaction terms are introduced and the system
is analyzed and solutions are given
at the classical level and at the quantum level by
using canonical quantization.
}
\vskip.5truecm
\vfill
\eject

\bigskip\centerline{\ut 1. Introduction}
\vskip1truecm
\TagsOnRight
During the last few years some progress was made towards
understanding and resolving the problems associated with
two-dimensional gravity.
(For a review see [1].)
A big effort has
been invested in this area after Polyakov revitalized the Liouville
action by quantizing it in the light-cone gauge [2,3]. The quantum theory
of the Liouville action was studied some time ago in [4] and
recently its appearance at the quantum level was understood from
first principles [5]. However, this approach of assuming the
dynamics of the metric to be governed by the Liouville action alone had
to stop at the so called $d=1$-barrier: quantization turns out
to be inconsistent if the conformal dimension of the coupled matter
is $1<d<25$ [2,3]. This seems to indicate that some
essential ingredient is missing.

Here, we want to promote the idea that the missing ingredient
is the dilaton. Actually, many classical (super-)gravity theories
in higher dimensions are known to be consistent only if a
dilaton $\phi$ and an antisymmetric
tensor field $b$ accompanies the metric
$g$. In two dimensions the field strength $H=db$ vanishes
and so here we need to consider only the dilaton.
Indeed, in [6,7] the dilaton has been seen to solve
the problem of two-dimensional gravity.
There the dilaton was introduced as a Langrange multiplier that
implies a constraint of constant curvature
and thus takes the Liouville
action on-shell.
Then quantization of string theory becomes
possible for any dimension $d$ and
the explicit result for the partition function could be given
on any genus [6].
The concept of constraining the system to constant curvature
has also been treated by Jackiw and Teitelboim [8].
The constraint of constant curvature, however, is {\it not essential}
for the dilaton solution of the Liouville problem. Here, we introduce
dilaton self-couplings that destroy its role
as a Lagrange multiplier. Nevertheless, the graviton-dilaton
coupling can be fixed in such a way that the quantum theory
remains free of the conformal anomaly.

Our main concern is to study the two-dimensional
graviton-dilaton-matter system [6-11] in the presence
of interaction terms both classically and quantum mechanically.
In section 2 we review aspects of the dilaton solution to the
Liouville problem and expose how dilaton self-couplings
have to be introduced.
In section 3 we consider the gravity action as a classical system
on its own right and give solutions.
In section 4 we include the Liouville contributions
and consider the full quantum system by canonical quantization.
We also give the Wheeler-De Witt equation and solutions.

\vskip1.2truecm

\vfill
\eject
\bigskip\centerline{\ut 2. The Dilaton Solution of the Liouville Problem}
\vskip0.8truecm
We begin with the classical action
$$
S(X,g,\phi) = S_g(g,\phi) + S_m(X,g)
\tag2.1
$$
Here $S_m$ describes the matter part of conformal dimension $d$
and $S_g$ is the gravity contribution
$$
S_g(g,\phi) = S_{EH}(g) + S_D(\phi,g)
\tag2.2
$$
with Einstein-Hilbert and dilaton action:
$$
\align
S_{EH}(g) &=
{1\over 2\pi}\int_M d^2x \sqrt{g} \left( \eta R_g + \mu\right)\tag2.3\\
S_{D}(\phi,g) &=
{1\over 2\pi}\int_M d^2x \sqrt{g} \left( b\phi R_g + \la e^\phi\right)
\tag2.4
\endalign
$$
where $\eta, b, \mu$ and $\la$ are constants.
The first term in (2.3) gives the
Euler number $\chi(M) = 2(1-h)$ for a closed and orientable manifold $M$
with $h$ handles, the second is the cosmological constant.
The first term in (2.4) is the usual coupling of the
dilaton $\phi$ to curvature, the last term introduces self-couplings.
Notice that with given normalization of the exponent
in (2.4) the $b$ can be fixed by quantum consistency.
Moreover, with self-coupling the dilaton
becomes non-trivial. Its role as a Lagrange multiplier
that was essential in [6,7] for computing the partition function
is lost. The dilaton will no longer trivialize
the Liouville problem. Nevertheless, the $b$ can be fixed such
that the self-coupling is of weight $(1,1)$ which implies
that the theory remains free of the conformal anomaly.
(For a discussion of including weight (1,1) fields
into an action see [12].)

Classically, a kinetic term for $\phi$ can easily be introduced
in (2.4) by rescaling $g_\ab\rightarrow e^\phi g_\ab$.
For the quantum theory, however, it is essential that
we quantize the gravity action in the form (2.4).
Otherwise the usual Liouville problems will re-appear.

The appearence of the Liouville action at the quantum level
is most easily seen from the path integral approach.
We work in the conformal gauge, where part of the
reparametrization invariance
is fixed by requiring
$$
g_{\alpha \beta}(x) = e^{2\sigma (x)}
{\hat g}({\tau})_{\alpha \beta}(x)
\tag2.5
$$
As usual the ${\hat g}$ is a representative in the conformal class
of $g$. These classes are specified by the Teichm\"uller parameter $\tau$
that will not be written in the following.
The field $\sigma$ is the Liouville mode and will be quantized
while $\hat g_{\alpha\beta}$ is a background field (see [1]).
A useful relation is
$$
R_g = e^{-2\sigma}\left(R_{\hat g} + \Delta_{\hat g}\sigma\right)
\tag2.6
$$
with $\Delta_{\hat g}={-1\over\sqrt{\hat g}}\pa_\al\sqrt{\hat g}
\hat g^{\al\be}\pa_\be$.

Going to the path integral we begin with the measure expressed
in terms of $g_\ab$. Pulling back this measure to $\hg_\ab$
introduces the Liouville action:
$$
\aligned
\Cal D_g\si\Cal D_g\phi\Cal D_gX&|det^{FP}_g|\,
\\&
=\,\,
\Cal D_{\hat g}\si\Cal D_\hg\Cal D_{\hat g}X|det^{FP}_{\hat g}|\,
e^{-{24-d\over 12}\, S_L(\si)}
\endaligned
\tag2.7
$$
where we included the Fadeev-Popov determinant arising
from the gauge fixing (2.5) and the Liouville action
$$
S_L(\sigma) = {1\over \pi} \int_M d^2x \sqrt{\hat g}
\left( {1\over 2}
{\hat g}^{\alpha\beta}\partial_{\alpha}\sigma\partial_{\beta}\sigma
+ \sigma R_{\hat g} + \kappa \exp {2\sigma} \right)
\tag2.8
$$
Without the dilaton the coefficient of $S_L$ in (2.7) would be
the familiar $-{25-d\over 12}$ of Liouville theory [5].
(If the Liouville mode is not quantized, the coefficient
is $-{26-d\over 12}$.)
The change of this coefficient in the presence of the
dilaton can be understood along the lines of [5].
The natural scalar product for a variation $\delta\phi(x)$ is
$$
||\delta\phi(x)||^2_g\,=\,
\int_M d^2x \sqrt{\hg} e^{2\si}\delta\phi\delta\phi
\tag2.9
$$
Therefore in the background metric $\hat g_{\ab}$
the $\Cal D_g\phi$ comes with the measure $e^{\si}\delta\phi$.
Going to $\Cal D_{\hat g}\phi$ with the measure $\delta\phi$ only
does then imply
$$
\Cal D_g\phi\,=\,\Cal D_{\hat g}\phi\,|\Cal J|
\tag2.10
$$
where $|\Cal J|$ is the determinant of
$$
\Cal J(x,y) =\,{e^{\si(x)}\delta\phi(x)\over\delta\phi(y)}
=\,e^{\si(x)}\delta^{(2)}(x-y)
\tag2.11
$$
The $|\Cal J|$ is the Jacobian of the transformation.
For to find its $\si$-dependence we look at the
variation $\si\rightarrow\si+\de\si$. Then
$$
\de \ln |\Cal J| = \de Tr \ln \Cal J = \int_M d^2x \de^{(2)}(0)\si(x)
\tag2.12
$$
We then need a regularization coming from the
heat kernel expansion:
$$
\aligned
\de^{(2)}(0)\,&=\, \sqrt g\left({1\over 4\pi\epsilon}
+ {1\over 12\pi}\,R_g\right)\\
&=\,\sqrt{\hat g}\left({e^{2\si}\over 4\pi\epsilon}
+ {1\over 12\pi}\,(R_{\hat g}+\Delta_{\hat g}\si)\right)
\endaligned\tag2.13
$$
with $\epsilon\rightarrow 0$. Substituting this into (2.12) implies
$$
|\Cal J|\,=\,e^{{1\over 12}S_L(\si)}
\tag2.14
$$
and this explains the shift in the Liouville coefficient
of (2.17). We see that the cosmological constant in $S_L$ is divergent.
The same holds for the contributions of the other measures
that can be calculated similarly to (2.9-14).
A counterterm has to be introduced that cancels this divergency.

Because of (2.7), the action (2.2) is accompanied
by the Liouville terms at the quantum level to give the effective action:
$$
\aligned
{1\over \al^{\prime}} S_g^{eff}(\ga\si,\ga^2\phi)
= &\,{1\over2\pi} \int_M d^2x \sqrt{\hat g}
      \Bigl({a}\hat g^{\al\be}\partial_\al\si\partial_\be\si
      + {Q\over \ga}\si R_{\hat g}
      + {1\over \ga^2} V(\ga\si,\ga^2\phi)\\ &\qquad\qquad
      + b(\ga\hat g^{\al\be}\partial_\al\phi\partial_\be\si
      +\phi R_{\hat g}) \Bigr)
      + {\eta\over \ga^2}\chi(M)
\endaligned\tag2.15
$$
where
$$
V(\ga\si,\ga^2\phi) = \mu e^{2\ga\si} + \la e^{2\ga\si+\ga^2\phi}
\tag2.16
$$
In (2.15) we kept the coefficients arbitrary.
We included the inverse string tension $\al^{\prime}\equiv\ga^2$
and rescaled $\si,\phi\rightarrow \ga\si, \ga^2\phi$.
Following the procedure of Kawai and Distler [3]
the coefficients in (2.15) can
be fixed by conformal field theory (CFT).
The energy momentum
tensor $\widehat T_{\alpha\beta}^g
= {-4\pi\over\sqrt{\hat g}}{\delta S\over\delta{\hat g}_{\alpha\beta}}$
that follows from (2.15) is
$$
\aligned
\widehat T_{\al\be}^g =\, &\de_{\al\be} ( {a}\pa^\ga\si\pa_\ga\si
               + b\ga\pa^\ga\phi\pa_\ga\si
               + {1\over\ga^2}V(\ga\si,\ga^2\phi)\, )\\
              &- 2a\pa_\al\si\pa_\be\si
               + {Q\over\ga}\left(\pa_\al\pa_\be\si
                     - \delta_{\al\be}\pa^2\si\right)\\
              &+ b\left(\pa_\al\pa_\be\phi
                     - \delta_{\al\be}\pa^2\phi
               - \ga\pa_\al\phi\pa_\be\si
                     - \ga\pa_\be\phi\pa_\al\si\right)
\endaligned
\tag2.17
$$
where we have chosen $\hg_\ab=\de_\ab$ on a local coordinate patch.
We first assume $V=0$ so that the $\si-\phi$ system is conformal.
Going to the coordinates $z={1\over \sqrt 2}(x^1 + ix^2),
\bar z={1\over \sqrt 2}(x^1 - ix^2)$ we get the operator product
expansion (OPE)
$$
\widehat T^g(z)\widehat T^g(w) = {{1\over 2}c_G\over (z-w)^4}
+ \left[ {2\over (z-w)^2} + {\pa_w\over z-w}\right] \widehat T(w) +...
\tag2.18
$$
where $\widehat T\equiv \widehat T_{zz}$ and
the conformal anomaly of the effective gravity system is
$$
c_G = 2 + {12\over\ga^2}\left( Q-a\right)
\tag2.19
$$
If we take the value $a$ from (2.7) and require the
anomaly (2.19) to cancel the anomaly $d-26$ arising from matter
and gauge fixing, we obtain
$$
Q\, =\, 2a\qquad,\qquad a = \al' {24-d\over 12}
\tag2.20
$$
Thus, we find the $\si R_\hg$ coefficient not to be renormalized.
This is already an improvement compared to Liouville theory
without dilaton since any renormalization of
the $\si R_\hg$-term destroys the geometrical character of the
theory.
Alternatively, we could have required this geometrical
content not to be broken at the quantum level and
(2.19) would have implied the $a$ value given in (2.20).

For $V\neq 0$ the conformal anomaly still vanishes if
the new terms are of weight $(1,1)$.
(For a discussion of this point see [12].)
We have to look at the OPE
$$
\widehat T(z):e^{\ga\si+\al\phi}: =
\left[ {h\over (z-w)^2} + {\pa_w\over z-w}\right]
:e^{\ga\si+\al\phi}:
\tag2.21
$$
where the conformal weight $h$ of the exponential is
$$
h = 1 + {\al\over b\ga}\left( {Q-2a\over 2\ga} - \ga
+ {a\over 2b\ga}\al\right)
\tag2.22
$$
Using (2.20) the conformal weight of $V$ will be (1,1) if we require
$$
b = \al'\,{24-d\over 24} = {a\over 2}
\tag2.23
$$
It is remarkable to see that the $\phi R_g$ coupling is of
order $\al'$, and thus a one-loop effect.
This is similar to the $\alpha^\prime$ expansion
of the effective string theory [13] where the $\phi R$ term is of
one order higher than the matter term.
The most important point to realize is that contrary to Liouville theory
without dilaton,
there is no gravitational dressing of the cosmological constant term
and (2.22) introduces no restriction on
the dimension $d$.

Without self-couplings, $\la=0$, the $\phi$ can be integrated out
and introduces the constraint of constant curvature.
Then topological consistency requires to replace
$$
R\,\rightarrow\,R\,+\,\Lambda
\tag2.24
$$
in (2.4) if $h\neq 1$ and we are back to the theory considered
in [6,7]. From the conformal field theory point
of view, the new term $\Lambda\phi e^{2\si}$ can be made a field
of weight $(1,1)$ if we also add $4\Lambda\si e^{2\si}$ [7].
Such a $\Lambda$-term describes a possible renormalization
after replacing (2.24), but
the $4\Lambda\si e^\si$ seems strange.
It can, however, arise from the linear part in $\Lambda$ of
the non-local action
$$
{a\over 2\pi}\int d^2x \sqrt{g} (R+\Lambda)\Delta^{-1}(R+\Lambda)
\tag2.25
$$
It is not clear to us whether such a combination does arise
in the non-perturbative calculation of the effective action.

It is possible to compare the action (2.2) with that of an effective
string theory of a two-dimensional target manifold [13], by rescaling
$$
g_\ab = \phi\, g^{\prime}_\ab\qquad ,\qquad \phi = e^{-2\varphi}
\tag2.26
$$
Then with $\la=0$ and replacement (2.24) in (2.4)
the action (2.2) takes the form
$$
S_g = {b\over 2\pi}\int_M d^2x\sqrt{g^{\prime}}
      \left[ e^{-2\varphi}\left(R^\prime
      + 2g^{\prime\ab}\pa_\al\varphi\pa_\be\varphi + {\mu\over b}\right)
      + \Lambda e^{-4\varphi}\right] + \eta \chi(M)
\tag2.27
$$
This is known to be the two-dimensional string
effective action with a dilaton
potential ${\mu\over b} e^{-2\varphi} + \Lambda e^{-4\varphi}$.
Apart from quantum consistency,
it should also be obvious why the rescaled action (2.27)
will not be considered. For example the $\la$-term in (2.4) would
become $\la\, \text{exp}(-2\varphi+\text{exp}(-2\varphi))$ !

If the coupled matter system is a string theory in
a d-dimensional Euclidean background:
$$
S_m(X,g) = -{1\over 8\pi}\int_M d^2x\sqrt{g}\,
g^{\al\be}\pa_\al X^i\pa_\be X^i
\tag2.28
$$
the $\si-\phi$ system leads to
a $D=d+2$ dimensional string theory with one time direction.
Absorbing the coefficient $a$ in $\si$ and $\phi$ gives back
the critical $D=26$ string in the limit $d\rightarrow 24$.

In summary, we have seen that (2.1) can be consistently
quantized for any $d$, provided that $b$ is given by (2.23).
We will now analyze the system (2.1) both at the classical
and the quantum level.

\vskip1.2truecm

\bigskip\centerline{\ut 3. The Classical Analysis}
\vskip0.8truecm
Before considering the full quantum theory we analyze
the system (2.1) as a classical system, i.e. without including
the Liouville terms induced from quantization.
Classically, the coefficient $b$ is arbitrary.
We rescale $\phi\rightarrow{1\over b}\phi$.
Also the case $\la=0$ will be discussed and so
we use the replacement (2.24). For definiteness
we take the matter action to be given by (2.28).

The variation of the action with respect to $g_\ab$ gives the
energy-momentum
tensor $T_\ab\equiv {-4\pi\over\sqrt{g}}{\delta S\over\delta g^\ab}$:
$$
\aligned
T_\ab = &\nabla_\al\nabla_\be\phi
             - \ {1\over 2}\pa_\al X^i\pa_\be X^i\\
        &+ g_\ab(-\nabla^\ga\nabla_\ga\phi + \Lambda\phi + \mu
                 +\lambda e^{\phi/b}
                 + {1\over 4}g^{\ga\de}\pa_\ga X^i\pa_\de X^i)
\endaligned
\tag3.1
$$
The $g_\ab, \phi$ and $X^i$ equations of motion are
$$
\align
T_\ab\,&= 0\tag3.2\\
R + \Lambda + {\lambda\over b}\,e^{\phi/b} &= 0\tag3.3\\
\pa_\al(\sqrt{g}g^\ab\pa_\be X^i) &= 0
\tag3.4
\endalign
$$
One interesting case occurs when $\lambda = 0$ and the genus of the
manifold $h\neq 1$. Then for consistency $\Lambda$ must be
different than zero and with a fixed sign.
We work in the conformal gauge (2.5)
where $\hg_\ab$ is a background metric chosen so that
$$
\alignat 3
\hg_{z\bar z}
&={2\over (1+z\bar z)^2}&\qquad &{\text if}\quad h=0,    &\, &R_\hg = 1\\
&=\,1                   &\qquad &{\text if}\quad h=1,    &\, &R_\hg = 0
   \tag3.5\\
&={-2\over (z-\bar z)^2}&\qquad &{\text if}\quad h\geq 2,&\, &R_\hg = -1
\endalignat
$$
If $h=0$ then $\Lambda<0$, and the constant curvature condition in (3.3)
has the solution
$$
e^{2\si} =
  -{1\over\Lambda}(1+z\bar z)^2 {f^\pr\bar f^\pr\over (1+f\bar f)^2}
\tag3.6
$$
A line element is then given by
$$
ds^2 =
  -{4\over\Lambda}{df\, d\bar f\over (1+f\bar f)^2}
\tag3.7
$$
and the conformal gauge can be fixed completely by
choosing $f(z)=z$, i.e. $e^{2\si}={-1\over\Lambda}$.
Similarly when $\Lambda>0,\, e^{2\si}={1\over\Lambda}$
and when $\Lambda=0,\, \si=0$.
The three components of $T_\ab$ in eq.(3.2)
(when $\Lambda < 0$) give
$$
\align
\pa\bar\pa\phi + {2\over (1+z\bar z)^2}(\phi + {\mu\over\Lambda}) &= 0\\
\pa^2 \phi +  {2\bar z\over (1+z\bar z)^2}\pa\phi
    - {1\over 2}\pa X^i\pa X^i &= 0\tag3.8\\
\bar\pa^2 \phi +  {2z\over (1+z\bar z)^2}\bar\pa\phi
    - {1\over 2}\bar\pa X^i\bar\pa X^i &= 0
\endalign
$$
The solution of the first equation involves an arbitrary
holomorphic function $u(z)$ and its conjugate $\bar u$ [14]:
$$
\phi = u(z) + {1-z\bar z\over 1+z\bar z}\int^z {dw\over w}\, u(w)
       - {\mu\over 2\Lambda} + c.c.
\tag3.9
$$
The solution in (3.9) must be subject to the constraints
imposed by the last two equations in (3.8), which imply
$$
\pa^2u + {\pa u\over z} - {u\over z^2} = {1\over 2}\pa X^i\pa X^i
\tag3.10
$$
and similarly for $\bar u$. Equation (3.4) simplifies to
$$
\pa\bar\pa X^i\, = \, 0
\tag3.11
$$
and the $X^i$ splits into holomorphic and antiholomorphic parts.
In the absence of matter ($X^i = 0$) the solution of (3.10)
can be found immediately to be
$$
u = {\al\over z} + \be z
\tag3.12
$$
and the dilaton field $\phi (z,\bar z)$ can be evaluated from (3.9)
to be
$$
\phi = -{\mu\over 2\Lambda} + {2\over 1+z\bar z}(-\al\bar z + \be z) + c.c.
\tag3.13
$$
In the presence of matter, the holomorphic part of $X^i$ is arbitrary,
and the $u(z)$ can be solved in function of it from eq (3.10).
The general solution is
$$
u(z) = {1\over c_1+c_2}(c_1u_1(z)+c_2u_2(z))
\tag3.14
$$
where
$$
\aligned
u_1(z) = {1\over 2}\int^z dz^\pr z^\pr
         \int^{z^\pr}dz^{\pr\pr} \pa X^i(z^{\pr\pr})\pa X^i(z^{\pr\pr})\\
u_2(z) = {1\over 2}\int^z dz^\pr z^{\pr\, 3}
         \int^{z^\pr}dz^{\pr\pr} z^{\pr\pr\, 2}
         \pa X^i(z^{\pr\pr})\pa X^i(z^{\pr\pr})
\endaligned
\tag3.15
$$
and $c_1$ and $c_2$ are arbitrary constants. To
compare with the much simpler case when $\Lambda = 0$,
the $T_\ab$ equations in (3.2) simplify to
$$
\align
\pa\bar\pa \phi &= \mu\\
\pa^2 \phi &= {1\over 2}\pa X^i\pa X^i\tag3.16\\
\bar\pa^2 \phi &= {1\over 2}\bar\pa X^i\bar\pa X^i
\endalign
$$
The solution of (3.16) is easily seen to be
$$
\aligned
\phi (z,\bar z) &= \mu z\bar z + g(z) +\bar g(z)\\
g(z) &= {1\over 2}\int^z dz^\pr\int^{z^\pr}dz^{\pr\pr}
        \pa X^i(z^{\pr\pr})\pa X^i(z^{\pr\pr})
\endaligned
\tag3.17
$$
In the absence of matter $(X^i=0)$, $g(z)$ simplifies to
$$
g(z) = d z + e
\tag3.18
$$
where $d$ and $e$ are arbitrary constants.
By shifting the coordinate $z$ by a constant, the dilaton $\phi$ can be put
into the form [11]
$$
\phi(z,\bar z) = \mu z\bar z + c
\tag3.19
$$
This can be seen to be the familiar black hole solution [15]
by recalling equation (2.26) where the metric of the
effective $2d$ string theory is $g^{\prime}_\ab=\phi^{-1}g_\ab$
making the line element
$$
ds^2 = {2dzd\bar z\over \mu z\bar z + c}
\tag3.20
$$
In this sense the solution in (3.13) can be considered as a
generalization of the black hole solution with non-trivial
topology. The case $\Lambda > 0$ corresponding to $h\geq 2$
can be treated along the same lines, and gives the solution
$$
\phi = u + \bar u - {\mu\over\Lambda} + {z+\bar z\over z-\bar z}
       \left[-\int^z {dz^\pr\over z^\pr}u
       + \int^z {d\bar z^\pr\over \bar z^\pr}\bar u\right]
\tag3.21
$$
where $u(z)$ satisfies eq. (3.10) and is solved by eqs. (3.14) and (3.15).

We now turn to the general case with $\lambda$ and $\Lambda\neq 0$.
The constant curvature constraint is removed and we do not have to
worry about the topological consistency. Notice that in this
case eq. (3.3) can be solved for $\phi$ in terms of $R$:
$$
\phi = - b
       \left[\ln(R+\Lambda) + \ln(-{b\over \lambda})\right]
\tag3.22
$$
and when this is substituted back into the action we get
for the gravity part
$$
-{b\over 2\pi}\int d^2x\sqrt{g}\left[(R+\Lambda)
  (1+\ln(-{b\over \lambda})+\ln(R+\Lambda))+{\mu\over b}\right]
\tag3.23
$$
In this form the gravitational action becomes completely
geometrical. In the conformal gauge it is a function of the
Liouville mode.
The comparison of the dynamical degress of freedom
in the actions (2.2) and (3.23) is amusing. In the first
there is only a mixed kinetic term of the form $\pa_\al\phi\pa^\al\si$.
Perturbatively it is a scalar field-ghost combination,
so the net degrees of freedom is zero.
In (3.23) only the field $\si$ appears and has a kinetic term.
However, if one expands perturbatively in $R$,
the propagator corresponds to ${1\over k^4}$ which can
be written
$$
{1\over k^4 + \epsilon^2} = {i\over 2\epsilon}
\left[{1\over k^2 + i\epsilon} - {1\over k^2 - i\epsilon}\right]
$$
The dynamical degree of this system is also zero.

We shall now solve the classical system (2.2) with (2.24).
There is no constraint of constant curvature.
First, we look for the solution of (3.2).
The full solution can be found exactly in the absence
of matter $(X^i=0)$.
If $\phi$ is constant the classical $g_\ab$ solutions
are of constant curvature.
If $\phi$ is not constant
then $\xi_\al={1\over\sqrt{g}}\epsilon_{\ab}\nabla^\be\phi$ is a
Killing vector [9] since (3.2) implies
that $\nabla_\al\nabla_\be\phi$ is proportional to
$g_\ab$ and therefore
$$
\align
\nabla_\al\xi_\be + \nabla_\be\xi_\al &= 0\tag3.24\\
\xi^\al\nabla_\al\phi &= 0\tag3.25
\endalign
$$
The Killing vector $\xi$ can be used to introduce new
coordinates $x, y$ where $y$ is in the direction of $\xi^\al$
and thus ignorable while $x$ is orthogonal,
with the line element:
$$
ds^2 = g_{xx}(x) dx^2 + g_{yy}(x) dy^2
\tag3.26
$$
In these coordinates the only non-vanishing Christoffel symbols
are $\Gamma_{xx}^x, \Gamma_{yy}^x$ and $\Gamma_{xy}^y$
and the $T_{xx}$ and $T_{yy}$ components of (3.2) give
$$
\aligned
g_{yy,x}\pa_x\phi &= 2 g_{xx} g_{yy} V(\phi)\\
g_{xx,x}\pa_x\phi &= 2 g_{xx} [\pa_x^2\phi - g_{xx} V'(\phi)]
\endaligned
\tag3.27
$$
where
$$
V'(\phi) = \Lambda\phi + \mu + \la e^{\phi/b}
\tag3.28
$$
while $T_{xy}$ is trivially zero.
Using the freedom of a coordinate
transformation $x\rightarrow x'(x)$,
we can choose
$$
g_{yy} = g^{xx} = h(x)
\tag3.29
$$
with this choice eqs. (3.27) simplify to
$$
\aligned
\pa_x^2\phi &= 0\\
\pa_xh\,\pa_x\phi &= 2V'(\phi)
\endaligned
\tag3.30
$$
and this can be immediately solved to give
$$
\aligned
\phi(x) &= A x + B\\
h(x) &= {2\over A}[\mu x + \Lambda({1\over 2}Ax^2 + Bx)
        +{\lambda \over A}e^{{1\over b}(Ax+B)}]
\endaligned
\tag3.31
$$
Moreover, the solution in (3.31) satisfies eq. (3.3)
since $R=-{1\over 2}\pa_x^2h$.

Let
$$
\aligned
z &= F(x)\, e^{iy}\\
\bar z &= F(x)\, e^{-iy}
\endaligned
\tag3.32
$$
The transformation function $F(x)$ in (3.32) and the
Liouville mode $\si (x)$ can be found from
$$
\aligned
ds^2 &= 2e^{2\si}dzd\bar z\\
     &= 2e^{2\si}[(\pa_x^2F)^2dx^2 + F^2dy^2]
\endaligned
\tag3.33
$$
Comparing with (3.26) and (3.29) we deduce that
$$
\aligned
F(x) &= F_0 e^{\pm\int^x{dx^\pr\over h(x^\pr)}}\\
e^{2\si(x)} &= {h(x)\over 2F^2(x)}
\endaligned
\tag3.34
$$
Obviously the inverse transformation from the (x,y) coordinates
to the $(z,\bar z)$ coordinates is very complicated.
The solutions in (3.31) and (3.34) are also a generalization
of the black hole solution in eq. (3.19).
It is remarkable that an exact solution for the
complicated geometrical theory in (2.2) can be found.
This would not have been possible without using the action
in the form (2.2) where the field $\phi (x)$ is not integrated out.

\vskip1.2truecm

\bigskip\centerline{\ut 4. The Quantum Analysis}
\vskip0.8truecm
We now turn to the quantum study of two-dimensional gravity
as given in (2.2). Quantum consistency requires $b$ to be given
by (2.23). The Liouville action has to be added and the
effective action turns out to be (2.15) with (2.20),(2.23):
$$
\aligned
S_g^{eff}(\si,\phi)
=\, &\eta\chi(M)+\al'{24-d\over 12\pi}\,\int_M d^2x \sqrt{\hat g}
   \Bigl({1\over 2}\hat g^{\al\be}\partial_\al\si\partial_\be\si\\
  &+ {1\over 4} \hat g^{\al\be}\partial_\al\phi\partial_\be\si
   + (\si+{\phi\over 4}) R_{\hat g}
   + {1\over 2} V(\si,\phi) \Bigr)
\endaligned
\tag4.1
$$
with
$$
V(\si,\phi) = \mu e^{2\si} + \la e^{2\si + \phi}
\tag4.2
$$
where we rescaled $\mu,\la\rightarrow a\mu,a\la$.
The action (4.1) reduces to the topological term in the
limit $\al'\rightarrow 0$. The $\phi$ and $\si$ equations
of motion are given by
$$
\align
\Delta_\hg\si + R_\hg + 2\la e^{2\si+\phi} &= 0\tag4.3\\
\Delta_\hg(\si + {\phi\over 4}) + R_\hg
  + \mu e^{2\si} + \la e^{2\si+\phi} &= 0\tag4.4
\endalign
$$

At this point the easiest approach to adopt is canonical
quantization.
The topology of $M$ is assumed to be that of a cylinder.
This allows the global choice of background metric
$$
\hg_\ab = \text{diag}(-1,1)
\tag4.5
$$
with $R_\hg = 0$. We shall follow the steps of Polchinski [16]
in his canonical quantization of the Liouville system.
Assuming the spatial coordinate $x^1\in [0,2\pi]$ is periodic
we can expand
$$
\align
\si(0,x^1) &= \si_0 - i {\sum_{n=-\infty}^\infty}^\pr
   {1\over n}(\al_ne^{inx^1}+\tilde\al_ne^{-inx^1})\\
\pa_+\si(0,x^1) &= \sum_{n=-\infty}^\infty
    \al_ne^{inx^1}\tag4.6\\
\pa_-\si(0,x^1) &= \sum_{n=-\infty}^\infty
    \tilde\al_ne^{-inx^1}
\endalign
$$
where $x^\pm = x^0\pm x^1$ and the prime on the
summation omits the $n=0$, and $\al_0=\tilde\al_0={1\over 2}p_\si$.
The analog expansion for the field $\phi$ is obtained
from the replacement $\al_n,\tilde\al_n\rightarrow\be_n,\tilde\be_n$
and $\be_0=\tilde\be_0={1\over 2}p_\phi$.

The canonical momenta are
$$
\aligned
\pi_\si \, &=\, -\, {a\over 4\pi}\,\pa_0\phi^\pr\\
\pi_{\phi^\pr} \, &=\, -\, {a\over 4\pi}\,\pa_0\si
\endaligned
\tag4.7
$$
where $\phi^\pr=\phi+2\si$ and $a$ was given in (2.20).
Then the usual quantization conditions on $\si, \pi_\si$
and $\phi^\pr,\pi_{\phi^\pr}$ result in the following commutators
$$
\aligned
{a\over 2}[\phi_0^\pr , p_\si ] &= i\\
{a\over 2}[\si_0,p_{\phi^\pr }] &= i\\
a [\al_n,\be^\pr_m] &= - n\,\de_{n+m,0} = a [\tilde\al_n,\tilde\be^\pr_m]
\endaligned
\tag4.8
$$

The energy-momentum tensor $\widehat T_\ab$ has been given
in (2.17) with (2.20),(2.23). The Virasoro
generators $L_n^{(\phi,\si)}$ and $\bar L_n^{(\phi,\si)}$
are defined as the Fourier components of $\widehat T_{++}+{c\over 24}$
and $\widehat T_{--}+{c\over 24}$. The shift of ${c\over 24}$
results from the fact
that $\widehat T_{++}$ does not transform as a tensor when the coodinates
are changed from the ones on the plane to
those on the cylinder.
Using the equations of motion in (2.17) to eliminate the
second $x^0$ derivatives, we find after some algebra
$$
\aligned
L_n^{(\phi,\si)}=-a\Big[&\sum_{k=-\infty}^\infty \al_{k}\be_{n-k}^\pr
  - i n(\al_n+\be_n^\pr)\\
  &+{1\over 4\pi}\int_0^{2\pi}dx^1 e^{-inx^1} V(\phi,\si)\Big]
  \quad,\quad n>0\\
L_0^{(\phi,\si)}=-a\Big[&\al_0\be_0^\pr+\sum_{k=1}^\infty
  (\al_{-k}\be_{k}^\pr+\be_{-k}^\pr\al_{k})\\
  &+{1\over 4\pi}\int_0^{2\pi}dx^1 V(\phi,\si)\Big] - {d-24\over 24}
\endaligned
\tag4.9
$$
and corresponding expressions for $\bar L_0^{(\phi,\si)}$
and $\bar L_n^{(\phi,\si)}$.

Including the matter and ghosts contributions, the quantum states are
$$
|\Psi> = |\Psi>_{g,m}|0>_{gh}
\tag4.10
$$
Then the BRST condition $Q|\Psi>=0$ implies
$$
\aligned
(L_n^{(\phi,\si)} + L_n^m)|\Psi>_{g,m} &= 0
= (\bar L_n^{(\phi,\si)} + \bar L_n^m)|\Psi>_{g,m}\, ,\quad n>0\\
(L_0^{(\phi,\si)} + L_0^m - 1)|\Psi>_{g,m} &= 0
= (\bar L_0^{(\phi,\si)} + \bar L_0^m - 1)|\Psi>_{g,m}
\endaligned
\tag4.11
$$
Assuming the matter fields to be in the highest weight states,
and the quantum state $|\Psi>_{g,m}$ factorize, then [16]
$$
\aligned
L_n^m|\Psi>_m &= \bar L_n^m|\Psi>_m = 0\, ,\quad n>0\\
L_0^m|\Psi>_m &= \bar L_0^m|\Psi>_m = \rho_m |\Psi>_m
\endaligned
\tag4.12
$$
where $\rho_M$ is some constant.

In the following we take $M$ to be 1+1 dimensional universe
where all fields are
assumed to be constant in the spatial direction $x^1$.
This is the mini-superspace approximation.
Before considering the quantum solutions it is
possible to consider the classical limit so the
only conditions to be satisfied are
$$
L_0^{(\phi,\si)} = \bar L_0^{(\phi,\si)}= 1 - \rho_m
\tag4.13
$$
In terms of the new variables $\Sigma=e^\si$ and $D=e^{\phi/2}$
the condition (4.13) reads
$$
a\, {\dot\Sigma\over\Sigma}\left({\dot\Sigma\over\Sigma}
  + {\dot D\over D}\right) =
   \rho_m - {d\over 24} - a \left( \mu + \la D^2 \right) \Sigma^2
\tag4.14
$$
Since the Liouville mode appears together with the
dilaton, eq. (4.14) is difficult to analyze.
If, however, we restrict to the classical solutions
so that $\Sigma, D$ have to obey (4.3), (4.4):
$$
\align
{1\over\Sigma^2}&\left[\Sigma\ddot\Sigma - (\dot\Sigma)^2\right]
   + 2\la\Sigma^2D^2 = 0\tag4.15\\
{1\over D^2}&\left[ D\ddot D - (\dot D)^2\right]
   + 4\mu\Sigma^2 - 4\la\Sigma^2 D^2 = 0
\tag4.16
\endalign
$$
the time derivative of (4.14) implies
$$
D^2\, = {\mu\over\la}
\tag4.17
$$
The condition (4.13) picks out the classical solution (4.17)
which is simple and of particular interest, since with (4.17)
the equations of motion (4.3), (4.4) reduce to the Liouville equation,
describing constant curvature on $M$:
$$
R_g\, = \, -\, 2\mu\tag4.18
$$
With (4.17) the condition (4.14) reduces to
$$
a\, \left({\dot\Sigma\over\Sigma}\right)^2 =
   \rho_m - {d\over 24} - 2 a\mu\Sigma^2
\tag4.19
$$
The $\Sigma$ equation of motion gives no further information since
it follows from (4.19). In (4.19) we recognize the
same structure of the expansion rate that was found
for pure Liouville theory [16], with the difference
that now the constant dilaton contributes to
the vacuum energy density. For
different values of $\rho_m, d$ and the cosmological
constant $\mu$ we find the theory to include
the de Sitter, Robertson-Walker, static and other
cosmological solutions that were discussed in [16].
Eq. (4.14) describes how these solutions are disturbed
in the presence of a varying dilaton,
if we allow for non-classical field configurations.

At the end of section two, we mentioned that $d\rightarrow 24$
gives back the critical $D=26$ string, where $D=d+2$.
This can also be seen from (4.14). If we let gravity decouple
by $\al'\rightarrow 0$ then $a\rightarrow 0$ and (4.14)
implies $\rho_m={d\over 24}$. Since matter without gravity
requires $\rho_m=1$ we find $d=24$ for the
critical case.

We now look at the quantum theory in the mini-superspace approximation
$$
\Psi(\si,\phi) = \Psi(\si_0,\phi_0)
\tag4.20
$$
We use (4.8) to replace
$$
\aligned
\al_0 &= {4\over ia}({1\over 2}{\pa\over\pa\si_0}+{\pa\over\pa\phi_0})\\
\be_0^{'} &= {4\over ia}{\pa\over\pa\phi_0}
\endaligned
\tag4.21
$$
Then the $L_0$ and $\bar L_0$ conditions imply the Wheeler-de Witt
equation:
$$
\left({\pa^2\over\pa\Sigma^2} + {1\over\Sigma}{\pa\over\pa\Sigma}
 + {D\over\Sigma}{\pa^2\over\pa\Sigma\pa D} - {a^2\over 16}
 (\mu + \lambda D^2) - {aM\over 16}{1\over\Sigma^2}\right)
 \Psi(\Sigma_0,D_0) = 0
\tag4.22
$$
where $M={d\over 24}-\rho_M$.

In this form (4.22) is a generalization of Bessel's equation.
Before, we have seen that the classical solution (4.17) covers
the dynamics of pure Liouville theory.
This can also be read from (4.22). If we restrict to the part
of $(\Sigma,D)$ space that is defined by (4.17) then
the $D$ independent solution of (4.22) turns out to
be of the same structure that was found in Liouville theory.
For $aM>0, \mu<0$ the eq. (4.22) is then solved by
$$
\Psi = \al J_{-n}({|a|\over 2}\sqrt{-(1/2)\mu}\,\Sigma)
      + \be J_{n}({|a|\over 2}\sqrt{-(1/2)\mu}\,\Sigma)
\tag4.23
$$
where $n={1\over 4}\sqrt{aM}$ and $\al, \be$ arbitrary.

In (2.9-14) the cosmological constant turned out to be
regularization dependent. We may use this freedom to set $\mu=0$.
In that case a $D$-dependent solution, defined on the
full $(\Sigma,D)$-space can be found for $aM>0,\la<0$:
$$
\Psi = \al J_{-m}({|a|\over 4}\sqrt{-(1/2)\la}\,\Sigma D)
      + \be J_{m}({|a|\over 4}\sqrt{-(1/2)\la}\,\Sigma D)
\tag4.24
$$
where $m={1\over 4}\sqrt{{1\over 2}aM}$.

Other solutions are obtained immediately if we set
the dilaton self-coupling to zero: $\la=0$.
Since the topology of $M$ is that of a cylinder,
this does not require the replacement (2.24).
Then for $\mu=0$ exponential expressions are found if we set
$$
\Psi = \Sigma^\al\,D^\be
\tag4.25
$$
This satisfies (4.24) if
$$
\al = -{\be\over 2} \pm {1\over 2}\sqrt{\be^2 + {aM\over 4}}
\tag4.26
$$
For $\be=0,\, aM<0$ eq.(4.26) includes the classically allowed
ingoing and outgoing solutions of Liouville theory.
For $\la=0,\,\mu\neq 0$ there are solutions independent
of the dilaton:
$$
\Psi = \al J_{-n}({|a|\over 4}\sqrt{-\mu}\,\Sigma)
      + \be J_{n}({|a|\over 4}\sqrt{-\mu}\,\Sigma)
\tag4.27
$$
which are of the type given in (4.23).

\vskip1.2truecm

\bigskip\centerline{\ut 5. Summary}
\vskip0.8truecm
In conclusion, we have shown that the two-dimensional
theory of gravity as described by a graviton-dilaton system
can be solved both at the classical and the quantum levels.
For the classical case, i.e. without including the Liouville
terms induced from quantization we found generalized
black hole solutions.
At the quantum level the induced Liouville action
has to be included.
In the absence of potential terms the theory is exactly conformal
and completely solvable.
In the presence of interactions,
the dilaton-graviton interaction
can be fixed in such a way that the theory remains
free of the conformal anomaly for any conformal dimension
of the coupled matter system.
We applied canonical quantization and gave the Wheeler-De Witt equation.
Although richer due to the presence of the dilaton,
the theory is seen to include the dynamics of pure Liouville theory.
{}From the two-dimensional cosmology point of view,
it allows for de Sitter, Robertson-Walker, static and
other cosmological solutions,
determined by the values of matter and
vacuum (cosmological constant plus dilaton) contributions.

\vskip1.5truecm

\bigskip
\centerline{\ut References}
\vskip0.8truecm
\item{[1]} E. D'Hoker,
{\sl Mod. Phys. Lett.\bf A 6} (1991) 745;
\newline
N. Seiberg,
Lectures given at
Cargese meeting on `Random surfaces, Quantum Gravity, and Strings',
{\sl Rutgers preprint} RU-90-29;

\item{[2]} A. M. Polyakov,
{\sl Mod. Phys. Lett. \bf A2} (1987) 893;

\item{[3]} V. G. Knizhnik, A. Polyakov and A. Zamolodchikov,
{\sl Mod. Phys. Lett.
\newline
\bf A3} (1988) 819;
\newline
A. H. Chamseddine and M. Reuter,
{\sl Nucl. Phys.
\newline
\bf B 317} (1989) 757;
\newline
F. David,
{\sl Mod. Phys. Lett. \bf A3} (1988) 1651;
\newline
J. Distler and H. Kawai,
{\sl Nucl. Phys \bf B 321} (1989) 509;

\item{[4]} T. L. Curtwright and C. B. Thorn,
{\sl Phys. Rev. Lett. \bf D 48} (1982) 1309;
\newline
E. Braaten, T. Curtwright and C. Thorn,
{\sl Phys. Lett.\bf B118} (1982) 115,
{\sl Ann. Phys. (NY) \bf 147} (1983) 365;
\newline
J. L. Gervais and A. Neveu,
{\sl Nucl. Phys \bf B 199} (1982) 59;

\item{[5]} N. Mavromatos and J. Miramontes,
{\sl Mod. Phys. Lett. \bf A4} (1989) 1847;
\newline
E. D'Hoker and P. S. Kurzepa,
{\sl Mod. Phys. Lett.\bf A 5} (1990) 1411;

\item{[6]} A. H. Chamseddine,
{\sl Phys. Lett.\bf B256} (1991) 379,
{\sl Phys. Lett.\bf B258} (1991) 97;

\item{[7]} A. H. Chamseddine,
{\sl Nucl. Phys \bf B 368} (1992) 98;

\item{[8]} R. Jackiw, $in$ Quantum Theory of Gravity,
ed. S. Christensen
(Adam Hilgers, Bristol, 1984), p. 403
\newline
C. Teitelboim, $in$ Quantum Theory of Gravity,
ed. S. Christensen
(Adam Hilgers, Bristol, 1984), p. 327;

\item{[9]} A. Tseytlin and C. Vafa,
{\sl Harvard preprint} HUTP-91/A 049;
\newline
A. Tseytlin,
{\sl Cambridge university preprint} DAMTP-37-1991;
\newline
J. Russo and A. Tseytlin ,
{\sl Stanford and Cambridge university preprint} SU-ITP-92-2, DAMTP-1-1992;

\item{[10]} S.D. Odintsov and I.L. Shapiro,
{\sl Phys. Lett. \bf B 263} (1991) 183;
{\sl Mod. Phys. Lett. \bf A 7} (1992) 437;
{\sl Madrid preprint} FTUAM 91-33.

\item{[11]} C. Callan, S. Gidding, J. Harvey and A. Strominger,
Santa Barbara preprint, UCSB-TH-91-54;

\item{[12]} A. Sen,
{\sl Nucl. Phys. \bf B345} (1990) 551;

\item{[13]} C. Callan, D. Friedan, E. Martinec and M. Perry,
{\sl Nucl. Phys \bf B 262} (1985) 593;
\newline
E. Fradkin and A. Tseytlin,
{\sl Nucl. Phys. \bf B 261} (1985) 1;

\item{[14]} A. H. Chamseddine and D. Wyler,
{\sl Nucl. Phys. \bf B 340} (1990) 595;

\item{[15]} M. Mueller,
{\sl Nucl. Phys. \bf B 337} (1990) 37;
\newline
E. Witten,
{\sl Phys. Rev. \bf D 44} (1991) 314;
\newline
S. Elitzur, A. Forge and E. Rabinovici,
{\sl Nucl. Phys. \bf B 359} (1991) 58;
\newline
G. Mandal, A. M. Sengupta and S. R. Wadia,
{\sl Mod. Phys. Lett.
\newline
\bf A6} (1991) 1685;

\item{[16]} J. Polchinski,
{\sl Nucl. Phys. \bf B 324} (1989) 123;

\end